\documentclass[a4paper]{jpconf}
\usepackage[latin1]{inputenc}
\usepackage{amsmath}
\usepackage{amsfonts}
\usepackage{amssymb}
\usepackage{listings}
\usepackage{graphicx}
\usepackage{epstopdf}

\usepackage{amsfonts,amssymb,amscd,amsmath}
\usepackage{amssymb}
\usepackage{amsmath}
\usepackage{graphicx}
\usepackage{graphics}
\usepackage{pifont}
\usepackage{hyperref}

\usepackage{enumitem}

\def\f12{\frac{1}{2}}

\usepackage{graphicx}
\begin{document}
\title{On the algorithm to find S-related Lie algebras}

\author{Carlos Inostroza$^{1}$, Igor Kondrashuk$^{2},$ Nelson Merino$^3$  \\ and Felip Nadal$^4$}

\address{$^1$  Departamento de F\'isica, Universidad de Concepci\'on, Casilla 160-C, Concepci\'on, Chile} 

\address{$^2$  Grupo de Matem\'atica Aplicada, Departamento de Ciencias B\'asicas, Univerdidad del B\'io-B\'io, Campus Fernando May, Casilla 447, Chill\'an, Chile}

\address{$^3$  APC, Universite Paris Diderot, 10, rue Alice Domon et Leonie Duquet, 75205 Paris Cedex 13, France}

\address{$^4$  Instituto de F\'isica Corpuscular (IFIC), Edificio Institutos de Investigaci\'on, \\
                c/ Catedr\'atico Jos\'e Beltr\'an, 2, E-46980 Paterna, Espa\~na}

\ead{nemerino@gmail.com}

\begin{abstract}
In this article we describe the Java library that we have recently constructed to automatize the S-expansion method, 
a powerful mathematical technique allowing to relate different Lie algebras.
An important input in this procedure is the use of abelian semigroups and thus, we start with a brief review about the classification of non-isomorphic semigroups made in the literature during the last decades, and explain how the lists of non-isomorphic semigroups up to order 6 can be used as inputs in many of the methods of our library.
After describing the main features of the classes that compose our library we present a new method called \textit{fillTemplate} which tuns out to be very useful to answer whether two given algebras can be S-related.
\end{abstract}

\section{Introduction}

The S-expansion \cite{irs} is a powerful mathematical technique allowing to relate different Lie algebras, which contains as a particular case the In\"{o}n\"{u}-Wigner contraction \cite{Segal,IW} 
and its generalizations \cite{WW3,hs,aipv1}. 
This method also turns out to be a valuable tool to study the relations between different physical theories. 
Since we already gave a technical description of the S-expansion in the related proceeding contribution \cite{PresNM}, as well as a review about its different physical applications, 
here we begin by summarizing the main ingredients needed to understand how the Java library \cite{webJava} works. We have constructed this library to automatize the S-expansion procedure.

Basically, the S-expansion method combines the structure constants of a Lie algebra with the inner multiplication law of an abelian semigroup in such a way that leads to the definition of other Lie algebras, 
called \textit{S-expanded} algebras. The inputs are:
\begin{itemize}
	\item a Lie algebra $\mathcal{G}$ with generators $\{X_{i}\}$ and Lie product $\left[  X_{i},X_{j}\right]  =C_{ij}^{k}X_{k}$ where $C_{ij}^{k}$ are the structure constants (convention sum on repeated indices is assumed).
	\item an abelian finite semigroup $S=\{\lambda_{\alpha},\alpha=1,\ldots,n\}$, whose inner operation can be represented either by its multiplication table 
	$A = (a_{\alpha \beta}) \equiv 	(\lambda_{\alpha} \cdot \lambda_{\beta} )$, with entries in $\lambda_{\alpha}$, or by the \textit{selectors} $K_{\alpha\beta}^{\rho}$ which 
	are defined through the relation $\lambda_{\alpha}\cdot\lambda_{\beta} =\lambda_{\gamma\left(  \alpha,\beta\right)  }=K_{\alpha\beta}^{\rho}\lambda_{\rho}$, 
	where $K_{\alpha\beta}^{\rho}=1$ if $\rho=\gamma\left(  \alpha,\beta\right)  $ and $K_{\alpha\beta}^{\rho}=0$ if $\rho\neq\gamma\left(  \alpha,\beta\right)  \,$. 
	Thus, the components of a selector can only take the values 0 or 1.
\end{itemize}

As is well-known, the structure constants and the selectors provide respectively a matrix representation for $\mathcal{G}$ and $S$. This allows to show that the Kronecker product of 
those representations  $\mathcal{G}_{S}$\ $\mathcal{=}$\ $S\otimes\mathcal{G}$ is also a Lie algebra with generators $X_{\left(  i,\alpha\right)  }\equiv\lambda_{\alpha}\otimes X_{i}$, Lie product 
defined by $\left[  X_{\left(  i,\alpha\right)  },X_{\left(  j,\beta\right)  }\right]
\equiv\lambda_{\alpha}\cdot\lambda_{\beta}\otimes\left[  X_{i},X_{j}\right]
=C_{\left(  i,\alpha\right)  \left(  j,\beta\right)  }^{\left(  k,\gamma
\right)  }X_{\left(  k,\gamma\right)  }\label{z3}$ and structure constants given by $C_{\left(  i,\alpha\right)  \left(j,\beta\right)  }^{\left(  k,\gamma\right)  }=K_{\alpha\beta}^{\gamma}C_{ij}^{k}\,$. 
The explicit proof can be found in Ref. \cite{irs}.

There are two cases where one can extract from $\mathcal{G}_{S}$ smaller algebras, with interesting properties. The first one happens when the original algebra has a subspace decomposition, 
e.g.\footnote{As formulated in the original work \cite{irs}, one can also deal with more general decompositions. However, on the first version of our library \cite{webJava} we only focus 
in this case $\mathcal{G}=V_{0}\oplus V_{1}$.}, of the type $\mathcal{G}=V_{0}\oplus V_{1}$ 
satisfying $\left[V_{0},V_{0}\right] \subset V_{0}\,, \ \left[  V_{0},V_{1}\right] \subset V_{1}\,,\ \left[  V_{1},V_{1}\right]  \subset V_{0}$. 
If the semigroup has a decomposition in subsets $S=S_{0}\cup S_{1}$ satisfying the so called \textit{resonant decomposition}: 
\begin{equation}
S_{0}\cdot S_{0}\in S_{0}\,,\ \ \ S_{0}\cdot S_{1}\in S_{1}\,,\ \ \ S_{1}\cdot
S_{1}\in S_{0}\label{r2}%
\end{equation}
then, one can show that $\mathcal{G}_{S,R} \equiv \left(  S_{0}\otimes V_{0}\right)  \oplus\left(S_{1}\otimes V_{1}\right)$ is a subalgebra of $\mathcal{G}_{S}$, which is called the \textit{resonant subalgebra}.
The other case occurs when the semigroup have a zero element $0_{S}$ satisfying $\lambda_{\alpha}%
\cdot0_{S}=0_{S}$ for any element $\lambda_{\alpha}\in S$. In that case the whole sector $0_{S}\otimes\mathcal{G}$ can be erased from $\mathcal{G}_{S}$ in such a way that what remains is still a Lie algebra, which is called the $0_{S}$\textit{-reduced algebra} $\mathcal{G}_{S,\rm red}$ (which is not necessarily a subalgebra of $\mathcal{G}_{S}$).

In summary, depending on the features of the abelian semigroup $S$ that one uses in the S-expansion, the following types of algebra can be generated
\begin{enumerate}[label=(\alph*)]
	\item the expanded algebra $\mathcal{G}_{S} = S \otimes \mathcal{G}$,
	\item the resonant subalgebra $\mathcal{G}_{S,R}$, if $S$ has at least one resonant decomposition of the type (\ref{r2}),
	\item the  $0_{S}$-reduced algebra $\mathcal{G}_{S,\rm red}$, if $S$ has a zero element  $0_{S}$,
	\item the  $0_{S}$-reduction of the resonant subalgebra $\mathcal{G}_{S,R,\rm red}$, if $S$ has both a zero element and a resonant decomposition.
\end{enumerate}

Having introduced the main ingredients used in the S-expansion method, in the next sections we give a description of the Java library \cite{webJava} that automatizes this procedure.
In Section 2 we review some aspects about the classification of non-isomorphic semigroups, while in Section 3 we briefly describe the library. 
Then, in Section 4 we explain how the lists of all non-isomorphic abelian semigroups can be used to classify all possible S-expanded algebras (a-d).
Finally, in Section 5, we will briefly describe a new method that extends our library and that is useful to answer if two given Lie algebras can be S-related. 

\section{Some aspects about the semigroup classification} 
As shown in the following table,
\begin{small}
\begin{equation}
\begin{tabular}
[c]{|l|l|l}\cline{1-2}%
order & $Q=\ $\# semigroups & \\\cline{1-2}%
2 & 4 & \\\cline{1-2}%
3 & 18 & \\\cline{1-2}%
4 & 126 & [Forsythe '54]\\\cline{1-2}%
5 & 1,160 & [Motzkin, Selfridge '55]\\\cline{1-2}%
6 & 15,973 & [Plemmons '66]\\\cline{1-2}%
7 & 836,021 & [Jurgensen, Wick '76]\\\cline{1-2}%
8 & 1,843,120,128 & [Satoh, Yama, Tokizawa '94]\\\cline{1-2}%
9 & 52,989,400,714,478 & [Distler, Kelsey, Mitchell '09]\\\cline{1-2}%
10 & \textbf{12,418,001,077,381,302,684} & [Distler, Jefferson, Kelsey,
Kotthoff '16]\\\cline{1-2}%
\end{tabular}
\ \label{hist}%
\end{equation}
\end{small}   
the problem of enumerating the all non-isomorphic finite semigroups of a certain order is a non-trivial problem, as the number $Q$ of semigroups increases very quickly with the order of the semigroup. 
This classification has been made by many different authors (see e.g. \cite{n4,n5,n6-3,n6-1,n7,n8,n9-1,n9-2,n9-3} and references therein). 
In lower orders, a list of the multiplication tables representing those semigroups can be explicitly constructed. 
For example, in Ref. \cite{Hildebrant} it has been claimed that its program \textit{gen.f} allows to generate, in lexicographical ordering, the lists \textit{sem.n} of these tables for the orders $n=2,...,7$. 
After running that program, we got them only up to order 6 (for some reason the program stops when reaches the 835,927th of the 836,021 semigroups of order 7). 
In any case, we remark that similar lists are not explicitly known for orders higher than 8 and that the number of non-isomorphic semigroups in those orders has been reached only 
by using indirect techniques (see e.g., \cite{n9-4,S10}).

As we will see, the lists \textit{sem.n} up to order $n=6$ can be used as inputs in our library (although its methods are not restricted to that order\footnote{The methods of our library also allow us to perform semigroup calculations (like checking associativity, commutativity, finding zero element, isomorphisms and resonances) when the order $n$ is higher than 6. The only issue is that we do not have the full list of non-isomorphic tables for those higher orders.}).
In each list, a semigroup $S_{\left(n\right)}^{a}$ of order $n$ is uniquely identified by the number $a=1,...,Q$ and the semigroup elements $\lambda_{\alpha}$ are represented by $\alpha$ with $\alpha=1,...,n$. 
The program \textit{com.f} of \cite{Hildebrant} selects from those lists only the abelian semigroups (which are the ones of interest in the context of the S-expansion). 
For example, for $n=2$ the semigroup elements are $\{1,2\}$ and the program \textit{com.f} gives:
\begin{equation}%
\begin{tabular}
[c]{l|ll}%
$S_{\left(  2\right)  }^{1}$ & $1$ & $2$\\\hline
$1$ & $1$ & $1$\\
$2$ & $1$ & $1$%
\end{tabular}
\ \ \ \ \text{, }%
\begin{tabular}
[c]{l|ll}%
$S_{\left(  2\right)  }^{2}$ & $1$ & $2$\\\hline
$1$ & $1$ & $1$\\
$2$ & $1$ & $2$%
\end{tabular}
\ \ \ \ \text{, }%
\begin{tabular}
[c]{l|ll}%
$S_{\left(  2\right)  }^{4}$ & $1$ & $2$\\\hline
$1$ & $1$ & $2$\\
$2$ & $2$ & $1$%
\end{tabular}
\,.
\label{list1}%
\end{equation}
However, to avoid renaming the abelian semigroups in a new list, in our library we use directly the lists \textit{sem.n} and select the abelian ones with a simple method when is needed\footnote{It is also worth to mention that the lists \textit{sem.n} are exhaustive, i.e., if one finds a semigroup $S$ of order $n\leq6$ whose multiplication table is not contained in the list \textit{sem.n}, then $S$ must be isomorphic to one and only one semigroup in that list. This can be easily checked with the methods implemented in our library and thus, we are able to work with arbitrary semigroups, i.e., not necessarily having a lexicographical ordering.}.

\section{Description of the Java library to perform S-expansions}

To use the Java library it is necessary to download the linear algebra package \textit{jama.jar}\footnote{In particular, we use the methods belonging to the class \textit{Matrix} of the library \textit{jama.jar}.} 
from \cite{jama} and the following files from \cite{webJava}: 
\begin{small}
\begin{equation}%
\begin{tabular}
[c]{|l|l|}\hline
\textbf{File} & \textbf{Brief description}\\\hline
\textit{data.zip} & Contains the the files \textit{sem.n}\\\hline
\textit{sexpansion.jar} & Is the library itself\\\hline
\textit{ReadMe.pdf} & File with detailed installation instructions\\\hline
\textit{examples.zip} & Example programs explained in Section 5 of the Ref. \cite{Inostroza:2017ezc}\\\hline
\textit{Output\_examples.zip} & Output samples of the example programs\\\hline
\end{tabular}
\ \label{table_files}%
\end{equation}
\end{small}
Our library is composed of 11 classes\footnote{To see the source code one only have to unzip the file \textit{sexpansion.jar}.}, each of them containing different methods allowing us to automatize the S-expansion procedure. 
As the full documentation of the library is available in \cite{wiki}, here we only describe the main features of these classes.

First, the class \textit{SetS} allows to represent a set of integers, where none of them is repeated. It contains methods allowing to represent permutations, which are useful to determine isomorphisms between different 
semigroup tables. It also contains methods allowing to generate all possible subsets $S_{0}$ and $S_{1}$ of a semigroup $S$, with respectively $n_{0}\leq n$ and $n_{1}\leq n$ elements, such that $S=S_{0}\cup S_{1}$ 
(notice that some elements can be simultaneously in $S_{0}$ and $S_{1}$). Thus, the methods of this class also provide the basics elements to the study of resonances.

The class \textit{Semigroup} is used to represent discrete semigroups by means of the multiplication table $A = (a_{\alpha \beta}) \equiv (\alpha \cdot \beta )$ and contains methods allowing us to perform basic semigroup 
operations like checking associativity, commutativity, finding the zero element and loading the lists \textit{sem.n} when is needed. 
It also contains methods that, when combined with those of the class \textit{SetS}, allows to calculate isomorphisms and study all possible resonant decompositions of any given semigroup.

The class \textit{Selector} allows to represent semigroups by means of their selectors of $K_{\alpha\beta}^{\rho}$ which, as explained in the introduction, are very convenient to describe S-expansions. It also contains methods allowing to determine if the semisimplicity is preserved under the S-expansion. The classes \textit{SelectorReduced}, \textit{SelectorResonant}, \textit{SelectorResonantReduced} are constructed as a child of the class \textit{Selector}, i.e., they extend that class for the cases where the semigroup have a zero element, a resonance and both simultaneously.

The class \textit{StructureConstantSet} allows to represent the original algebra $\mathcal{G}$ by means of its structure constants $C_{ij}^{k}$. It also contains methods to perform basics calculations, like to determine the Killing-Cartan metric of $\mathcal{G}$.

The class \textit{StructureConstantSetExpanded} allows to represent the Lie algebra resulting of performing the S-expansion of $\mathcal{G}$ with $S$.
Finally, the child classes \textit{StructureConstantSetExpandedReduced}, \textit{StructureConstantSetExpandedResonant} and \textit{StructureConstantSetExpandedResonantReduced} allow to represent the expanded algebras $\mathcal{G}_{S,\rm red}$, $\mathcal{G}_{S,R}$ and $\mathcal{G}_{S,R,\rm red}$ when it corresponds.

Further details can be found in Ref. \cite{Inostroza:2017ezc}, which has been written as a handbook to use the library. Indeed, it explains in detail how the methods of our library has been constructed, 
how do they work and also explains the example programs that allow to check the results obtained in Refs. \cite{Caroca:2011qs} and \cite{Andrianopoli:2013ooa}.

\section{Classification of all possible S-expansions}

The first motivation that lead us to construct our library was to provide the computational tool to study all possible S-expansions of the type (a-d) that one can perform with all the abelian non-isomorphic semigroups 
provided by the lists \textit{sem.n}. On one side, this implies to identify all the abelian semigroups having a zero element, something that can be easily done not only by our library but also by the programs 
given in \cite{Hildebrant}. On the other side, a study of all possible resonant decompositions of these semigroups is also needed. To our knowledge, this was not done in the literature before 
our works \cite{Caroca:2011qs} and \cite{Andrianopoli:2013ooa} where we elaborated the first basic methods to study resonances, which we have improved recently and presented in the form of 
a Java library \cite{webJava} in Ref. \cite{Inostroza:2017ezc}. 
Thus, this is the first computational tool allowing to perform a full study of those resonances, as well as to classify and represent all possible S-expansions of the type (a-d) of a given Lie algebra. 
This result can be summarized in the following table,
\begin{small}
\begin{equation}%
\begin{tabular}
[c]{|l|c|c|c|c|c|}\hline
& $n=2$ & $n=3$ & $n=4$ & $n=5$ & $n=6$\\\hline%
\begin{tabular}
[c]{|l|}\hline
\#$\mathcal{G}_{S}$\\\hline
\#{\small pss}\\\hline
\end{tabular}
& \multicolumn{1}{|r|}{%
\begin{tabular}
[c]{|l|}\hline
$3$\\\hline
$2$\\\hline
\end{tabular}
} & \multicolumn{1}{|r|}{%
\begin{tabular}
[c]{|r|}\hline
$12$\\\hline
$5$\\\hline
\end{tabular}
} & \multicolumn{1}{|r|}{%
\begin{tabular}
[c]{|l|}\hline
$58$\\\hline
$16$\\\hline
\end{tabular}
} & \multicolumn{1}{|r|}{%
\begin{tabular}
[c]{|l|}\hline
$325$\\\hline
\multicolumn{1}{|r|}{$51$}\\\hline
\end{tabular}
} & \multicolumn{1}{|r|}{%
\begin{tabular}
[c]{|l|}\hline
$2,143$\\\hline
\multicolumn{1}{|r|}{$201$}\\\hline
\end{tabular}
}\\\hline%
\begin{tabular}
[c]{|l|}\hline
\#$\mathcal{G}_{S,\text{red}}$\\\hline
\#{\small pss}\\\hline
\end{tabular}
& \multicolumn{1}{|r|}{%
\begin{tabular}
[c]{|l|}\hline
$2$\\\hline
$1$\\\hline
\end{tabular}
} & \multicolumn{1}{|r|}{%
\begin{tabular}
[c]{|l|}\hline
$8\,$\\\hline
$3$\\\hline
\end{tabular}
} & \multicolumn{1}{|r|}{%
\begin{tabular}
[c]{|l|}\hline
$39\,$\\\hline
\multicolumn{1}{|r|}{$9$}\\\hline
\end{tabular}
} & \multicolumn{1}{|r|}{%
\begin{tabular}
[c]{|l|}\hline
$226\,$\\\hline
\multicolumn{1}{|r|}{$34$}\\\hline
\end{tabular}
} & \multicolumn{1}{|r|}{%
\begin{tabular}
[c]{|l|}\hline
$1,538\,$\\\hline
\multicolumn{1}{|r|}{$135$}\\\hline
\end{tabular}
}\\\hline%
\begin{tabular}
[c]{|l|}\hline
\#$\mathcal{G}_{S,R}$\\\hline
\#r\\\hline
\#{\small pss}\\\hline
\end{tabular}
& \multicolumn{1}{|r|}{%
\begin{tabular}
[c]{|l|}\hline
$1$\\\hline
$1$\\\hline
$1$\\\hline
\end{tabular}
} & \multicolumn{1}{|r|}{%
\begin{tabular}
[c]{|l|}\hline
$8$\\\hline
$9$\\\hline
$1$\\\hline
\end{tabular}
} & \multicolumn{1}{|r|}{%
\begin{tabular}
[c]{|r|}\hline
$48$\\\hline
\multicolumn{1}{|l|}{$124$}\\\hline
$4$\\\hline
\end{tabular}
} & \multicolumn{1}{|r|}{%
\begin{tabular}
[c]{|r|}\hline
$299$\\\hline
\multicolumn{1}{|l|}{$1,653$}\\\hline
$7$\\\hline
\end{tabular}
} & \multicolumn{1}{|r|}{%
\begin{tabular}
[c]{|r|}\hline
$2,059\,$\\\hline
\multicolumn{1}{|l|}{$25,512$}\\\hline
$23$\\\hline
\end{tabular}
}\\\hline%
\begin{tabular}
[c]{|l|}\hline
\#$\mathcal{G}_{S,R,\text{red}}$\\\hline
\#r\\\hline
\#{\small pss}\\\hline
\end{tabular}
& \multicolumn{1}{|r|}{%
\begin{tabular}
[c]{|l|}\hline
$0$\\\hline
$0$\\\hline
$0$\\\hline
\end{tabular}
} & \multicolumn{1}{|r|}{%
\begin{tabular}
[c]{|l|}\hline
$5$\\\hline
$6$\\\hline
$1$\\\hline
\end{tabular}
} & \multicolumn{1}{|r|}{%
\begin{tabular}
[c]{|l|}\hline
$32$\\\hline
$92$\\\hline
\multicolumn{1}{|r|}{$1$}\\\hline
\end{tabular}
} & \multicolumn{1}{|r|}{%
\begin{tabular}
[c]{|r|}\hline
$204$\\\hline
\multicolumn{1}{|l|}{$1,295$}\\\hline
$6$\\\hline
\end{tabular}
} & \multicolumn{1}{|r|}{%
\begin{tabular}
[c]{|r|}\hline
$1,465$\\\hline
\multicolumn{1}{|l|}{$20,680$}\\\hline
$12$\\\hline
\end{tabular}
}\\\hline
\end{tabular}
\ \ \ \ \ \ \label{cp_table1}%
\end{equation}
\end{small}
It gives for each order the number of S-expansions of the type (a-d) that can be performed, as well as the number $\#pss$ of S-expansions preserving semisimpliciy. 
Notice, a given semigroup can have more than one resonance, and with them the same semigroup may lead to different resonant subalgebras. 
Thus, in the table we count not only the number of abelian semigroups having at least one resonance, but also the total number of different resonances $\#r$.

\section{A method to find S-related algebras}

Imagine that we have to find a semigroup with some multiplications fixed, for example
\begin{equation}%
\begin{tabular}
[c]{l|llll}%
& $\lambda_{1}$ & $\lambda_{2}$ & $\lambda_{3}$ & $\lambda_{4}%
$\\\hline
$\lambda_{1}$ &  & $\lambda_{3}$ &  & $\lambda_{4}$\\
$\lambda_{2}$ & $\lambda_{3}$ &  & $\lambda_{4}$ & $\lambda_{4}$\\
$\lambda_{3}$ &  & $\lambda_{4}$ &  & $\lambda_{4}$\\
$\lambda_{4}$ & $\lambda_{4}$ & $\lambda_{4}$ & $\lambda_{4}$ & $\lambda_{4}$%
\end{tabular}\,.%
\end{equation}
As in Java an empty space cannot be left in an array, we choose to represent
them by  $-1$ in its place. For example, the template above will be
represented by%
\begin{equation}
\left(
\begin{tabular}
[c]{rrrr}%
$-1$ & $3$ & $-1$ & $4$\\
$3$ & $-1$ & $4$ & $4$\\
$-1$ & $4$ & $-1$ & $4$\\
$4$ & $4$ & $4$ & $4$%
\end{tabular}
\right)  \,.\label{ex_template}%
\end{equation}
Clearly, there are $4^{4}=256 $ different ways to fill this template in a commutative way, where only some of them are
associative so they are really semigroups. 
In general, for a template of order $n$ where $x$ independent component has been fixed, there are
$n^{n\left(  n+1\right)  /2-x}$ different forms to fill it in a commutative way. 
This task is performed by the method \textit{fillTemplate} of the class \textit{Semigroup}, which reads the \textit{template} and returns a \textit{list} with all the tables that fill it. It is used as follow.
\begin{small}
\begin{lstlisting}[language=Java]
list = Semigroup.fillTemplate(template);
\end{lstlisting}
\end{small}
From all those tables one should select those which  are associative and satisfy, if apply, a certain resonance.
Applying the isomorphisms methods, we can finally identify the set of non-isomorphic semigroups that are solutions for the given template. 
As it will be explained in \cite{Inostroza}, this method turns out very useful to elaborate a general algorithm to answer if two given Lie algebras can be S-related.

\section{Final Remarks}

As our library has an open licence GNU, we expect that it can be improved and extended, for example to study other kind of decompositions like, e.g., $S=S_{0}\cup S_{1}\cup S_{2}$ where
\begin{align*}
S_{0}\times S_{0}  & \subset S_{0}\ ,\ \ S_{0}\times S_{1}\subset
S_{1}\ ,\ \ S_{0}\times S_{2}\subset S_{2}\\
S_{1}\times S_{1}  & \subset S_{0}\cap S_{2}\ ,\ \ S_{1}\times S_{2}\subset
S_{1}\ ,\ \ S_{2}\times S_{2}\subset S_{0}\cap S_{2} \,.%
\end{align*}
This could be useful for new physical applications because, as shown in \cite{irs}, this type of decomposition can be used to study
expansions of Lie super algebras (used in the context of supergravity and string theory), which have the subspace decomposition $\mathcal{G}$ $=V_{0}\oplus V_{1}\oplus V_{2}$ with the following structure
\begin{align*}
\left[  V_{0},V_{0}\right]    & \subset V_{0}\ ,\ \ \left[  V_{0}%
,V_{1}\right]  \subset V_{1}\ ,\ \ \left[  V_{0},V_{2}\right]  \subset V_{2}\\
\left[  V_{1},V_{1}\right]    & \subset V_{0}\oplus V_{2}\ ,\ \ \left[
V_{1},V_{2}\right]  \subset V_{1}\ ,\ \ \left[  V_{2},V_{2}\right]  \subset
V_{0}\oplus V_{2}\,.%
\end{align*}

On the other hand, as mentioned in \cite{Nesterenko2012} it it would be interesting to analyze if the S-expansion could help to fit to the classification of solvable Lie algebras of a fixed dimension using S-expansions of semisimple Lie algebras of the same dimension.
Thus, our library might also be useful to analyze that problem.

\ack{We thank Andrei Kataev for his invitation to present this work at ACAT 2017. C.I. was supported by a Mecesup PhD grant and the T\'ermino de tesis grant from CONICYT (Chile). C.I. is very grateful to Local Organizing Committee and in particular to Gordon Watts for the financing his participation in ACAT.  
I.K. was supported by was supported by Fondecyt (Chile) grant 1050512 and by DIUBB (Chile) Grant Nos. 102609 and GI 153209/C. 
NM is supported by a Becas-Chile postdoctoral grant.}

\end{document}